\documentclass[pra,superscriptaddress,twocolumn]{revtex4-1}
\usepackage[utf8]{inputenc}
\usepackage[english]{babel}
\usepackage{amsmath}
\usepackage{amsfonts}
\usepackage{amssymb}
\usepackage{graphicx}
\usepackage{amsthm}
\usepackage{color}
\usepackage{hyperref}
\usepackage[caption=false]{subfig}
\newcommand{\ket}[1]{\ensuremath{|#1\rangle}}
\newcommand{\bra}[1]{\ensuremath{\langle #1|}}

\newcommand{\ketbra}[1]{\ensuremath{| #1 \rangle\langle #1 |}}
\newcommand{\dketbra}[2]{\ensuremath{| #1 \rangle\langle #2 |}}
\newcommand{\norm}[1]{\ensuremath{\left|\left|#1\right|\right|}}
\newcommand{\eqq}[1]{Eq.~(\ref{#1})}
\newcommand{\tr}[2]{\ensuremath{\operatorname{Tr}_{#1}\left[#2\right]}}
\newcommand{\abs}[1]{\ensuremath{\left|#1\right|}}

\newcommand{\ave}[1]{\ensuremath{\left\langle#1\right\rangle}}

\begin{document}
\title{Asymmetric information capacities of reciprocal pairs of quantum channels} 
\author{Matteo Rosati}
\affiliation{F\'{\i}sica Te\`{o}rica: Informaci\'{o} i Fen\`{o}mens Qu\`{a}ntics, Departament de F\'{\i}sica, Universitat Aut\`{o}noma de Barcelona, 08193 Bellaterra, Spain}
\affiliation{NEST, Scuola Normale Superiore and Istituto Nanoscienze-CNR, I-56127 Pisa, Italy.}
\author{Vittorio Giovannetti} 
\affiliation{NEST, Scuola Normale Superiore and Istituto Nanoscienze-CNR, I-56127 Pisa, Italy.}

\begin{abstract}  
Reciprocal pairs of quantum channels  are defined as completely positive
transformations which  admit a rigid, distance-preserving, yet not completely-positive transformation that allows to reproduce the outcome of one from the corresponding outcome of the other. 
From a classical perspective these transmission lines 
should exhibit the same communication efficiency. This is no longer the case in the quantum setting: explicit 
asymmetric behaviours are reported studying the classical communication capacities of reciprocal pairs of depolarizing and Weyl-covariant channels. 
\end{abstract}
\maketitle

\section{Introduction}\label{sec:intro}
Quantum mechanics is well-known for giving rise to counterintuitive effects with respect to the classical world we experience. Quantum information theory often relies on harnessing such effects to produce technological applications. One such effect that has attracted much attention in the past arises when trying to communicate a chosen direction to a distant party~\cite{peresFlip,massarFlip,gisinFlip,munozTapiaFlip,massarFlip2}. Indeed it turns out that, if we encode the directional information on the state of two quantum spin systems, such direction can be more efficiently estimated when using antiparallel spins rather than parallel ones~\cite{gisinFlip,munozTapiaFlip,massarFlip2}. This asymmetry is quite counterintuitive from a classical point of view and it is caused by two peculiarities of quantum mechanics: i) the impossibility of perfectly inverting the direction of an unknown spin; ii) the possibility of performing entangling measurements that introduce stronger-than-classical correlations between the two spins. 

\begin{figure}[t!]
\includegraphics[scale=.30]{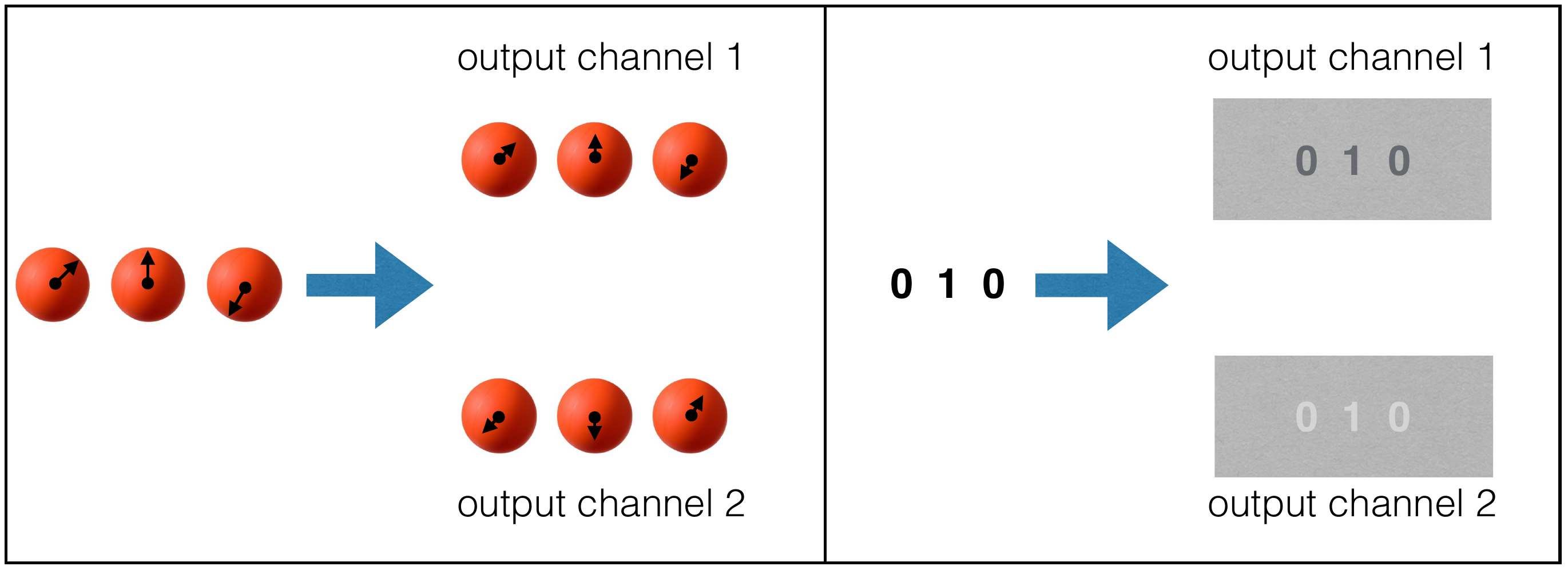}
\caption{Pictorial representation of the action of two  qubit depolarizing channels with opposite values of the depolarizing parameters: channel 1 simply shrinks the spin while channel 2 also reverse its orientation.
The action of these maps can alternatively being view as adding/subtracting the input signal from a completely noisy background (right panel). }\label{figure1new}
\end{figure}

In this article we report the presence of similar asymmetries in the broader context of classical communication on noisy quantum channels.
We start by considering the qubit depolarizing channel (DC)~\cite{nChuangBOOK,depolarizingCCap}, whose well-known action in phase-space is that of shrinking the Bloch vector of a qubit state by a given factor $\lambda\leq1$ that can take negative values. Hence, we can consider pairs of qubit DCs with parameters $\pm\abs{\lambda}$, whose output states are attenuated versions of the inputs but with opposite directions in the Bloch 
sphere, see left panel of Fig.~\ref{figure1new}.
Classical intuition suggests that 
these two channels should exhibit identical information-transmission efficiency. However, as for the direction-transmission problem cited above, this is not the case: when exploiting entanglement as a side resource to boost the communication process~\cite{bennetEntAss1,bennetEntAss2} one can send information at a higher rate using the qubit DC that shrinks and inverts the input rather than the one that just shrinks it by the same amount. 
   This asymmetry is perhaps even more striking than the one discussed in Refs.~\cite{peresFlip,massarFlip,gisinFlip,munozTapiaFlip,massarFlip2}, since in the latter case  improvements arise only when the sender
   of the message
   alternates inverted and non-inverted signals, while in the present case the noise-induced inversion acts on each exchanged signal. 
 
A generalization of this result to higher-dimensional DCs can be obtained by exploiting the fact that all 
these maps (including the qubit case)  can be  described as noisy processes that 
 take the input state of the transmitted signal and, depending on the sign of the parameter $\lambda$,
add it  to, or subtract it from a completely mixed background, 
see right panel  Fig.~\ref{figure1new}.
Once more, from a purely classical point of view  the two procedures are expected to possess the same 
 information-transmission capabilities. Yet it turns out  that the same asymmetry observed for the qubit case also applies for systems of higher dimensionality. 
Furthermore, at variance with the two-dimensional case, this effect is observed  also when quantum side resources are absent, i.e.,  when considering 
 the unassisted classical capacity~\cite{holevo1,holevo2,holevo3,schumawest1,schumawest2} of the channels.   
Eventually,  we further generalize our findings by introducing pairs of reciprocal maps $\{\mathbf{D}^{(\Phi)}_{\pm|\lambda|}\}$  obtained by adding to/subtracting from  the completely mixed state a proper fraction of the outputs of an arbitrary channel $\Phi$. Similarly to the DC case, these maps 
generate output images of the set of input density matrices which can be mapped one into the other via a  rigid (distance preserving) transformation. Once more, while from a classical point of view these channels should exhibit  identical communication performances, in the quantum case an asymmetric behaviour can be
observed. In particular we show this effect by considering the case where $\Phi$ belongs 
 to the class of finite-dimensional Weyl-covariant channels (WCCs)~\cite{holWeylCov,holevoRemarksFull,holDattaWCC} for which we compute the explicit value of the
  entanglement-assisted capacity. 

The article is structured as follows: in Sec.~\ref{sec:qubitDep} we discuss the basic asymmetric property of DCs; in Sec.~\ref{SECREC} we introduce the notion of generalized reciprocal pairs. The special case of  WCCs is addressed in Sec.~\ref{sec:weylCov} where 
the associated entanglement-assisted capacity is computed for the first time;  in Sec.~\ref{sec:conc} we compare the results with the spin case and discuss future lines of research. 
The paper ends with a series of technical appendices. 

\section{Capacity asymmetries for reciprocal DC pairs}\label{sec:qubitDep}

In quantum information theory,  
 the most general  physical transformations that a quantum system $S$ can undergo when travelling through a noisy medium
 are described in terms of quantum channels, i.e.,  
 linear, completely positive and trace-preserving maps $\Phi$   operating on the space $\mathfrak{S} ({\cal H})$ of the density matrices $\hat{\rho}$ of $S$~\cite{HOLEVOREV,wildeBOOK}. A benchmark of the communication efficiency attainable when employing $S$ as a carrier
 of classical information between two distant parties  is the classical (unassisted) capacity $C_{UA}(\Phi)$ of the channel, e.g., the highest transmission rate achievable when using $\Phi$ asymptotically many times with possibly entangled inputs and measurements~\cite{holevo1,holevo2,holevo3,schumawest1,schumawest2}. When shared entanglement between the sender and the receiver of the messages is available,  the proper figure of merit  is instead the entanglement-assisted classical capacity $C_{EA}(\Phi)$~\cite{bennetEntAss1,bennetEntAss2} which accounts for the possibility of exploiting the super-coding effect~\cite{SUPER} in boosting
the communication efficiency of the channel. 

In the particular case of a two-dimensional quantum system, a qubit, the density matrices $\hat{\rho}$ of $S$ admit a one-to-one parametrization in terms
of 3-dimensional Bloch vectors $\vec{r}$ in the unit sphere of $\mathbb{R}^3$,
\begin{equation}
\hat{\rho}=\frac{\hat{\mathbf{1}}+\vec{r}\cdot\vec{\hat{\sigma}}}{2},
\end{equation}
 with $\hat{\sigma}_{j}$, $j=1,2,3$, being the ordinary  spin-$1/2$ Pauli matrices while  $\hat{\mathbf{1}}$ being the identity operator on $\mathcal{H}_{2}$~\cite{nChuangBOOK}. 
 In this context 
 a qubit depolarizing channel $\mathcal{D}^{(2)}_{\lambda}$ can be defined as a mapping that induces an isotropic contraction of the Bloch  sphere
 by rescaling all its elements by an amount $\lambda$ (the depolarizing parameter) \begin{equation}
\mathcal{D}^{(2)}_{\lambda}:\vec{r}\mapsto\lambda\vec{r}.\label{QUBIT}
\end{equation}
While simple geometrical considerations suggest that not all real values of  $\lambda$ are admissible,
non-positive values of this parameter are  allowed.  As a matter of fact, a study of the complete-positivity
of the transformation~(\ref{QUBIT}) reveals that it defines a suitable quantum channel if and only if 
  $\lambda$ lays in the interval $[-1/3,1]$~\cite{ruskaiQubitMaps,depolarizingCCap,kingUnital}. 
  Accordingly, for each $\abs{\lambda}\leq 1/3$ we can construct reciprocal pairs of qubit DCs  with opposite values of the depolarizing parameter, i.e.,
  $\{\mathcal{D}^{(2)}_{\pm\abs{\lambda}}\}$. When applied to the same input state, the two elements of each of such pairs yield output Bloch vectors that have  equal norm, uniformly shrunk by $|\lambda|$ with respect to the initial value $\abs{\vec{r}}$, but opposite verses, with  $\mathcal{D}^{(2)}_{-\abs{\lambda}}$ 
  inducing an effective, orientation-independent inversion on the qubit state.  \\
 Going beyond the qubit setting, 
 the same construction can also be realized 
   when $S$ is a generic $d$-dimensional quantum system. 
\begin{figure}[t]
\includegraphics[scale=.45]{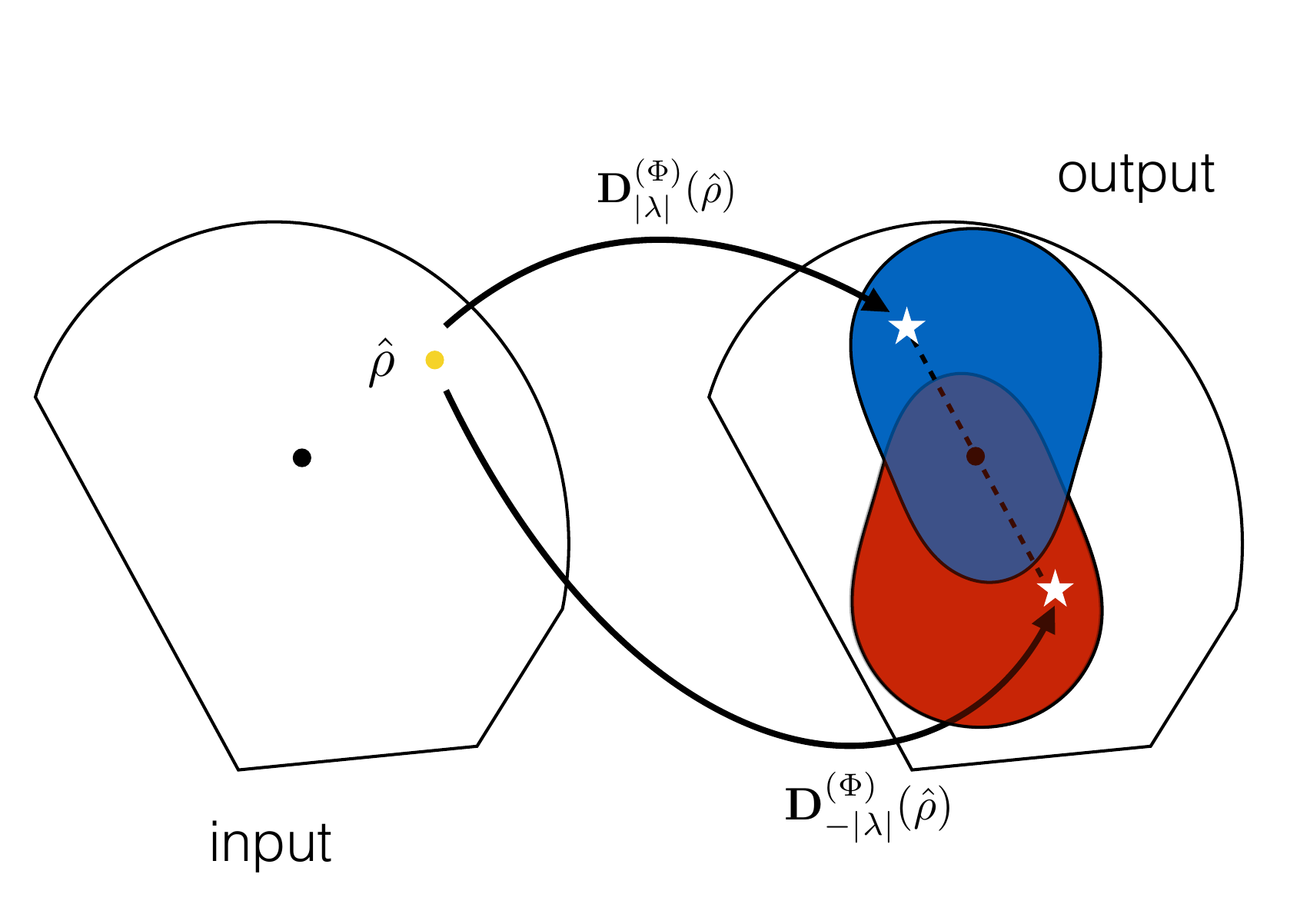}
\caption{Graphical representation of the action on the density matrix space $\mathfrak{S} ({\cal H})$ of the reciprocal pairs $\mathbf{D}^{(\Phi)}_{\pm|\lambda|}$ of Eq.~(\ref{dcGEN}) associated with the quantum channel $\Phi$: the blue and red sets represent the  output images associated with $\mathbf{D}^{(\Phi)}_{|\lambda|}$ and 
$\mathbf{D}^{(\Phi)}_{-|\lambda|}$ respectively,  which are symmetric under inversion by its central point ${\hat{\mathbf{1}}}/{d}$ 
 (black dot in the figure). The reciprocal DCs  $\{ \mathcal{D}^{(d)}_{\pm |\lambda|}\}$  are obtained by setting $\Phi$ equal to the identity channel ${\rm Id}$.  }\label{figurenew2}
\end{figure}
 Here DCs  can be  introduced as the following linear combination between
 the identity channel, ${\rm Id}: \hat{\rho} \rightarrow \hat{\rho}$, and the completely depolarizing channel, $\mathcal{D}_0^{(d)}$,
 that sends all inputs into the completely mixed state ${\hat{\mathbf{1}}}/{d}$~\cite{depolarizingCCap}:
\begin{equation}\label{dc}
\mathcal{D}_{\lambda}^{(d)} =\lambda {\rm Id} + \left(1-\lambda\right) \mathcal{D}_0^{(d)}  \;,
\end{equation}
which is equivalent to Eq.~(\ref{QUBIT}) in the special case $d=2$.\\
 DCs possess several interesting properties:
 \begin{enumerate}
 \item They are unital, i.e., the completely mixed state is a fixed point under the action of these maps, $\mathcal{D}_{\lambda}^{(d)}(\mathbf{1}/d)=\mathbf{1}/d$;
\item They are the only CP maps that 
commute with the action of the unitary group in any dimension~\cite{uNot3,twirling1};  
\item They form a convex semigroup under composition, i.e., for all $p\in[0,1]$
\begin{eqnarray}
\mathcal{D}_{\lambda_1}^{(d)}\circ \mathcal{D}_{\lambda_2}^{(d)} &=&\mathcal{D}_{\lambda_1 \lambda_2}^{(d)}\;, \label{SEMIG}
 \\ \label{CONVEX} 
p \mathcal{D}_{\lambda_1}^{(d)} +  (1-p) \mathcal{D}_{\lambda_2}^{(d)} &=&\mathcal{D}_{p \lambda_1 + (1-p) \lambda_2}^{(d)}\;; 
\end{eqnarray} 
\item  They are optimal quantum movers~\cite{qMovers}, meaning that 
the density matrices $\mathcal{D}_{\lambda}^{(d)}\left(\ketbra{\psi}\right)$ they produce when applied to pure states have
a fixed input-output fidelity~\cite{nChuangBOOK}  value, i.e.,
\begin{eqnarray} \label{MOVERS} 
\bra{\psi}\mathcal{D}_{\lambda}^{(d)}\left(\ketbra{\psi}\right)\ket{\psi} = f(\lambda):= \frac{1+(d-1)\lambda}{d}\;,
\end{eqnarray} 
$\forall |\psi\rangle \in {\cal H}$, whose minimum is optimal among all channels, see Appendix~\ref{movers} for details.
\end{enumerate}
As in the qubit case, only restricted values of the depolarizing parameter 
$\lambda$ permit to identify~(\ref{dc}) as a proper quantum channel. In generic dimension $d$ these are given by the real interval $[\lambda_{m}(d),1]$, with 
\begin{eqnarray} \lambda_{m}(d)=-1/(d^{2}-1)\; \end{eqnarray} being a negative quantity~\cite{depolarizingCCap}.  
Therefore, in full similarity with the qubit case, for each 
 $|\lambda|\leq |\lambda_{m}(d)|$
we can again define reciprocal pairs of 
DCs   $\{ \mathcal{D}_{\pm |\lambda|}^{(d)}\}$ with opposite depolarization coefficients. 
It is easily verified  that  both $\mathcal{D}_{|\lambda|}^{(d)}$ and $\mathcal{D}_{-|\lambda|}^{(d)}$ induce uniform contractions on the set $\mathfrak{S} ({\cal H})$ of density matrices toward its center, identified with the completely 
mixed state $\hat{\mathbf{1}}/{d}$. Indeed the latter is a fixed point of the maps, while the distance between any two output states is given by 
\begin{eqnarray}
 \norm{\mathcal{D}_{\pm |\lambda|}^{(d)}\left(\hat{\rho}_1\right)-\mathcal{D}_{\pm |\lambda|}^{(d)}\left(\hat{\rho}_2\right)} &=&  \label{DISTalso} 
|\lambda|\;  \norm{\hat{\rho}_1-\hat{\rho}_2}\;,
\end{eqnarray} 
where $\| \cdots \|$ is an arbitrary operator norm and $\hat{\rho}_1,\hat{\rho}_2 \in \mathfrak{S} ({\cal H})$ are generic input states, following from Eq.~(\ref{dc}).
Note that Eq.~(\ref{DISTalso}) also applies when $\{\mathcal{D}_{\pm |\lambda|}^{(d)}\}$ operate locally on joint states of $S$ with an arbitrary ancillary system $A$.  
Yet, while $\mathcal{D}_{ |\lambda|}^{(d)}$ operates by adding a fraction of the input state $\hat{\rho}$ to $\hat{\mathbf{1}}/{d}$, 
 $\mathcal{D}_{- |\lambda|}^{(d)}$ does the same by 
 subtracting the former from the latter. This realizes an effective inversion of
 the initial state $\hat{\rho}\in\mathfrak{S} ({\cal H})$, such that the output states of each reciprocal pair are found at the same distance from the completely mixed state but on opposite sides of it, see Fig.~\ref{figurenew2}, i.e.,
 \begin{eqnarray}
  \norm{ \mathcal{D}_{|\lambda|}^{(d)}\left(\hat{\rho}\right)-\mathcal{D}_{-|\lambda|}^{(d)}\left(\hat{\rho}\right)} &=&   
2 \norm{ \mathcal{D}_{\pm|\lambda|}^{(d)}\left(\hat{\rho}\right)-\frac{\mathbf{1}}{d}}\;.
 \end{eqnarray}
At variance with the scheme analyzed 
   in Refs.~\cite{peresFlip,massarFlip,gisinFlip,munozTapiaFlip,massarFlip2}, the extra flipping introduced  by $\mathcal{D}_{- |\lambda|}^{(d)}$
   is not part of a communication strategy established by the two 
  signaling parties. Instead
   it is automatically  induced by the environmental noise affecting the transmission process. It makes sense to ask whether this mechanism
 could  bring some advantages in terms of communication efficiency, since it assigns lower values of the input-ouput fidelity~(\ref{MOVERS}) to the negative component of a reciprocal pair, i.e., $f(-|\lambda|) \leq f(|\lambda|)$. 
 As a preliminary observation we consider the tensor-product channel obtained by two uses of the DC, possibly with different values of the depolarizing parameter: $\mathcal{D}^{(d)}_{\lambda}\otimes\mathcal{D}^{(d)}_{\lambda'}$. For a fixed value of $\abs{\lambda}=\abs{\lambda'}$, we would then expect to obtain a higher information-transmission rate in the case of opposite channel parameters, i.e., $\lambda=-\lambda'$, than that of equal ones, i.e., $\lambda=\lambda'$, in a similar fashion to the case addressed in Refs.~\cite{peresFlip,massarFlip,gisinFlip,munozTapiaFlip,massarFlip2}.
 We notice however that 
   both the unassisted classical capacity $C_{UA}$ and the entanglement-assisted capacity $C_{EA}$ of DCs are additive under tensor product with any other channel~\cite{depolarizingCCap,bennetEntAss2}, i.e.,
  $C_\ell (\mathcal{D}^{(d)}_{\lambda}\otimes\mathcal{D}^{(d)}_{\lambda'})=C_\ell (\mathcal{D}^{(d)}_{\lambda})+C_\ell(\mathcal{D}^{(d)}_{\lambda'})$ the label $\ell$ referring to the unassisted case (UA) or the entanglement-assisted one (EA).
  Therefore any asymmetry effect arising in this setting
  is not imputable to the joint use of channels with opposite depolarizing parameters and 
   cannot be directly linked to the case studied in Refs.~\cite{peresFlip,massarFlip,gisinFlip,munozTapiaFlip,massarFlip2}:
 if present it  can only be ascribed to 
  individual properties of the  positive and negative components of a reciprocal pair. 
   \begin{figure}[t!]
\includegraphics[scale=.45]{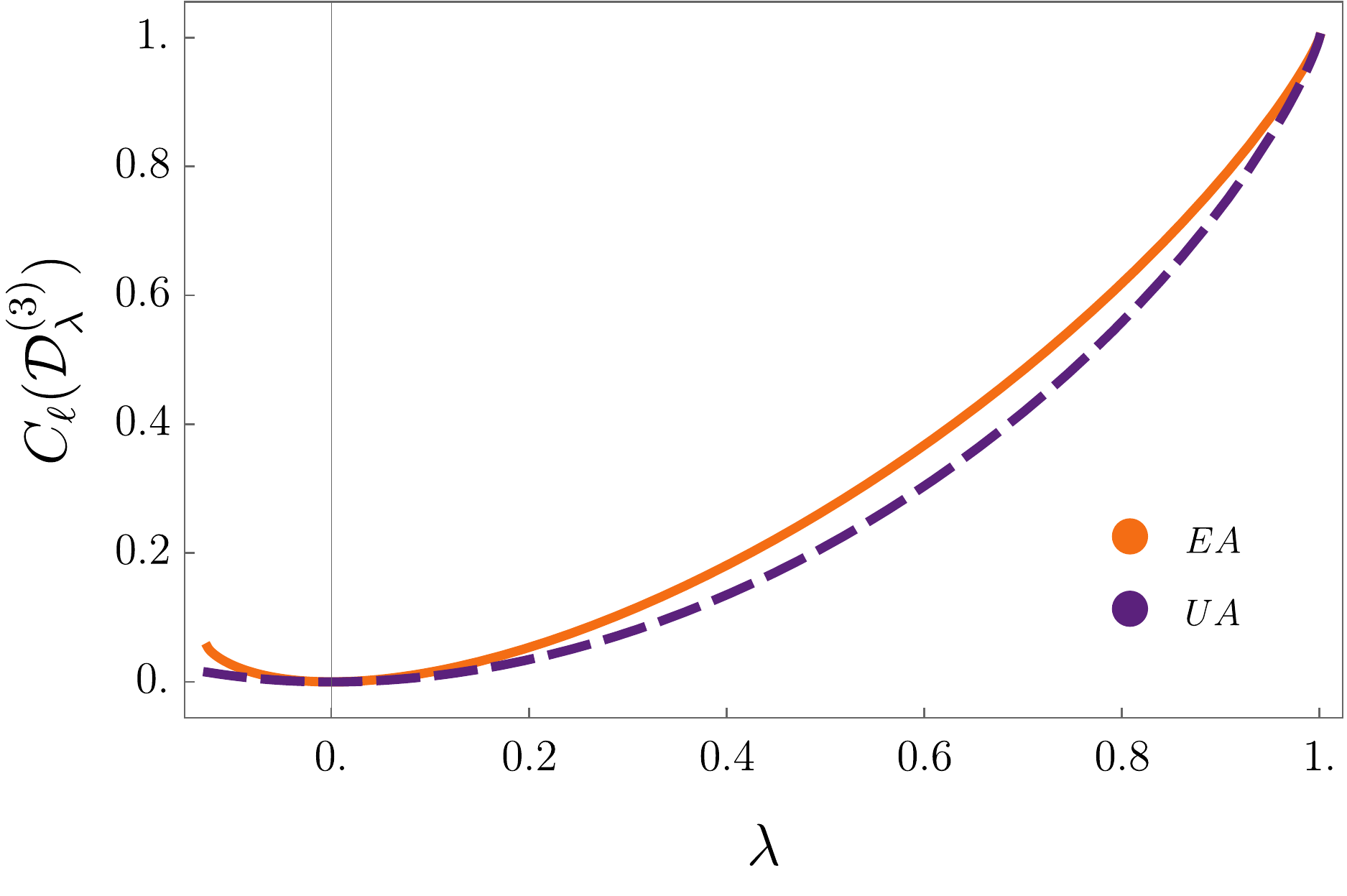}\put(-192.5,88.5){\includegraphics[scale=.22]{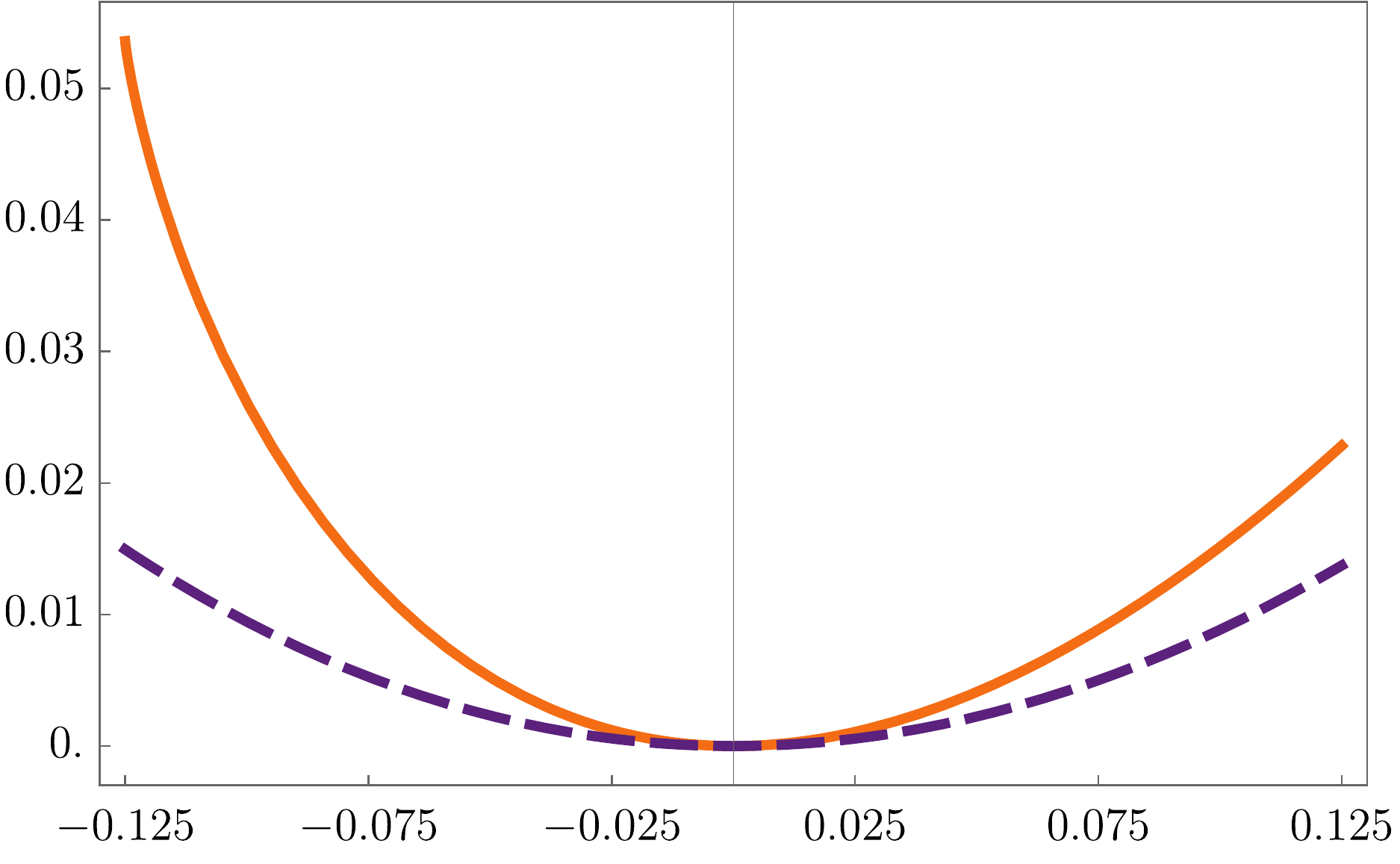}}
\caption{Plot of the  UA (purple dashed)  and the EA (orange solid) capacities of the DC for $d=3$, Eqs.~(\ref{CAPACITA},\ref{CAPACITAEA}), as a function of the  depolarizing parameter, $\lambda\in[\lambda_{m}(3),1]$. In the region where reciprocal pairs of DCs, $\{\mathcal{D}^{(3)}_{\pm\abs{\lambda}}\}$, can be defined (see the inset), the attained values of both capacities are higher for the negative-parameter channel of the pair, exhibiting a counterintuitive asymmetry. The effect is stronger in the EA case.}
\label{figSingle}
\end{figure} 
In order to check this, it is sufficient to recall  Refs.~\cite{depolarizingCCap,bennetEntAss1}, where 
 closed expressions for the UA and EA classical capacity of an arbitrary DC, ${\cal D}_\lambda^{(d)}$, 
 were provided in terms of the minimal value attained by the von Neumann entropy~\cite{nChuangBOOK} at the output of the map, $S_{min}\left(\mathcal{D}_{\lambda}^{(d)}\right)$.
 Specifically we have 
 \begin{eqnarray}
&C_{UA}\left(\mathcal{D}_{\lambda}^{(d)}\right)=\log d-S_{min}\left(\mathcal{D}_{\lambda}^{(d)}\right), \label{CAPACITA}\\
&C_{EA}\left(\mathcal{D}_{\lambda}^{(d)}\right)=C_{UA}\left(\mathcal{D}_{\lambda}^{(d^{2})}\right), \label{CAPACITAEA} 
\end{eqnarray}
and \begin{equation}\small{
S_{min}\left(\mathcal{D}_{\lambda}^{(d)}\right)=H\left(\left\{\frac{1+(d-1)\lambda}{d},\left(\frac{1-\lambda}{d}\right)^{\times (d-1)}\right\}\right)},
\end{equation}
the rhs of the last equation being the Shannon entropy of a probability distribution with $d-1$ equal terms
 and one equal to $\frac{1+(d-1)\lambda}{d}$ (all the logarithms are computed in base 2). 
 It turns out that for all $d\geq 3$, in the interval where flipping of the sign of the depolarizing parameter  is allowed, i.e., $|\lambda|\leq |\lambda_{m}(d)|$,
   both (\ref{CAPACITA}) and (\ref{CAPACITAEA}) are not symmetric: the capacities associated with the negative components of the reciprocal pairs are always larger than those of their positive counterparts, i.e.,
 \begin{eqnarray} C_{\ell}\left(\mathcal{D}_{-|\lambda|}^{(d)}\right) \geq C_\ell \left(\mathcal{D}_{|\lambda|}^{(d)}\right)\;,\label{ASS} \end{eqnarray} 
 see Fig.~\ref{figSingle}. For $d=2$ instead, the asymmetry is present only for the EA capacity and not for the UA one, suggesting that quantum correlations are strictly necessary to activate this property in the qubit case.
It is worth noticing that the above results cannot be associated with the convex semigroup properties~(\ref{SEMIG},\ref{CONVEX}), which imply the following inequalities for the capacities:
  \begin{eqnarray} &&C_{\ell}\left(\mathcal{D}_{\lambda_1\lambda_2}^{(d)}\right) \leq \min \left\{  C_\ell \left(\mathcal{D}_{\lambda_1}^{(d)}\right)\;,
  C_\ell \left(\mathcal{D}_{\lambda_2}^{(d)}\right)\right\}  \label{AASS}\;, \\
 &&C_{\ell}\left(\mathcal{D}_{\bar{\lambda}}^{(d)}\right) \leq p C_\ell\left( \mathcal{D}_{\lambda_1}^{(d)}\right)+  (1-p) C_\ell\left(\mathcal{D}_{\lambda_2}^{(d)}\right) \;,\nonumber
  \end{eqnarray}
   for all $\lambda_1$, $\lambda_2 \in [\lambda_m(d),1]$ and $p\in[0,1]$, where the first relation is derived via the data-processing inequality~\cite{HOLEVOREV}.\\
 Fig.~\ref{fig2} reports a detailed analysis of the observed capacity gaps between the elements of a reciprocal pair by plotting the 
asymmetry parameters 
\begin{equation}\label{asyBothDC}
a_{\ell}(\abs{\lambda})=\frac{C_{\ell}\left(\mathcal{D}_{-\abs{\lambda}}^{(d)}\right)-C_{\ell}\left(\mathcal{D}_{\abs{\lambda}}^{(d)}\right)}{C_{\ell}\left(\mathcal{D}_{\abs{\lambda}}^{(d)}\right)}
\end{equation}
as a function of the depolarizing parameter $|\lambda|$, for several system dimensions. Note that the EA asymmetry is about one order of magnitude stronger than the UA one and it is present also for $d=2$. Moreover, for a fixed value of $\abs{\lambda}$, higher asymmetry ratios are attained at higher dimensions. For fixed dimension instead, the maximum asymmetry ratio is attained at the edge of the allowed values of depolarization, i.e., $\abs{\lambda}=\abs{\lambda_{m}(d)}$, corresponding to the least noisy channel that still has a reciprocal one. This maximum, $a_{\ell}(\abs{\lambda_{m}(d)})$, is dimension-dependent and its behaviour is different depending on which capacity we are computing: $a_{EA}(\abs{\lambda_{m}(d)})$ increases with the dimension, saturating to a value $~\sim0.59$, while $a_{UA}(\abs{\lambda_{m}(d)})$ peaks at a value $\sim0.094$ for $d=4$ and then decreases to zero.
\begin{figure}[t!]
\subfloat[]{\includegraphics[scale=.45]{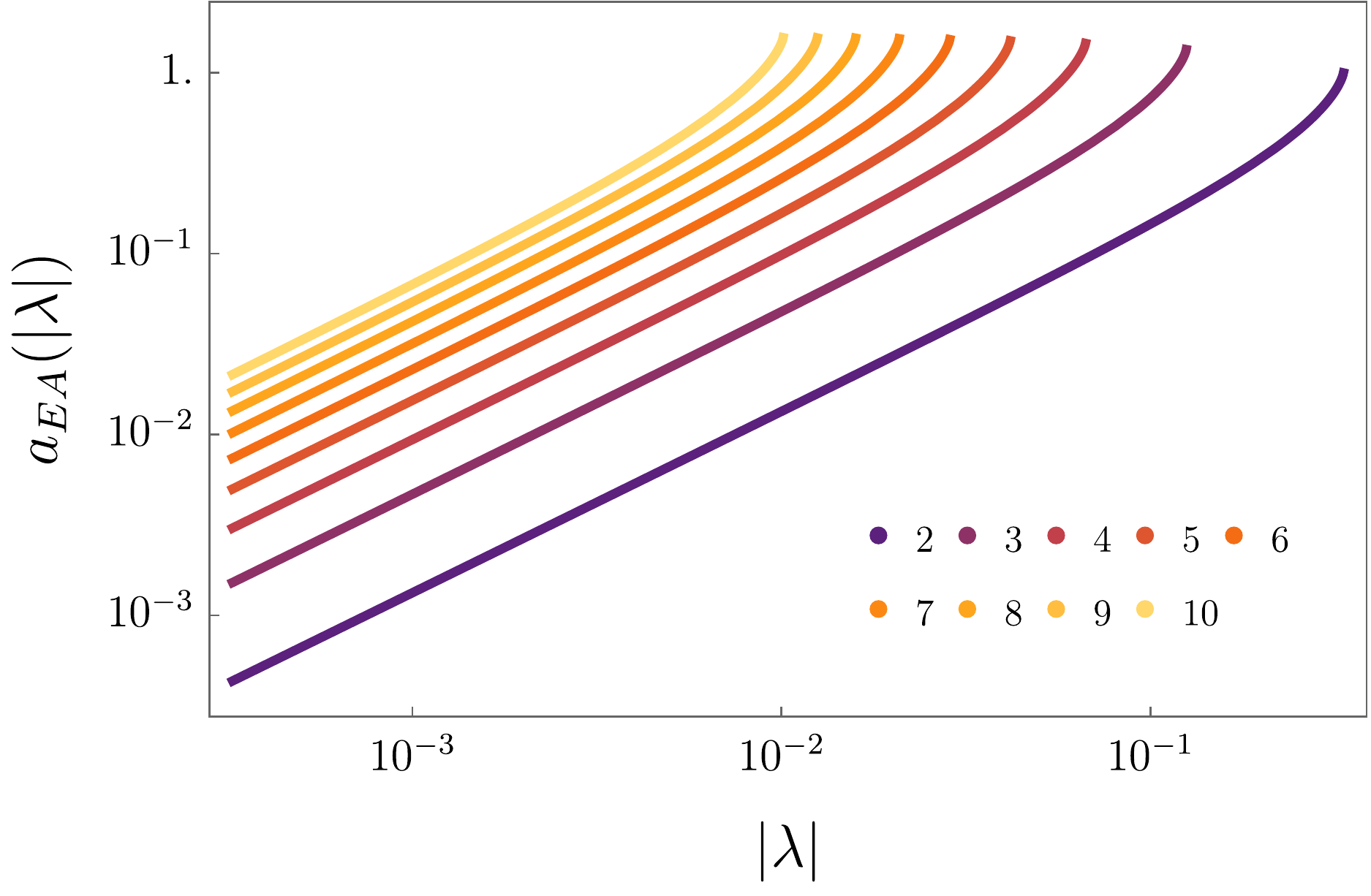}}\\
\subfloat[]{\includegraphics[scale=.45]{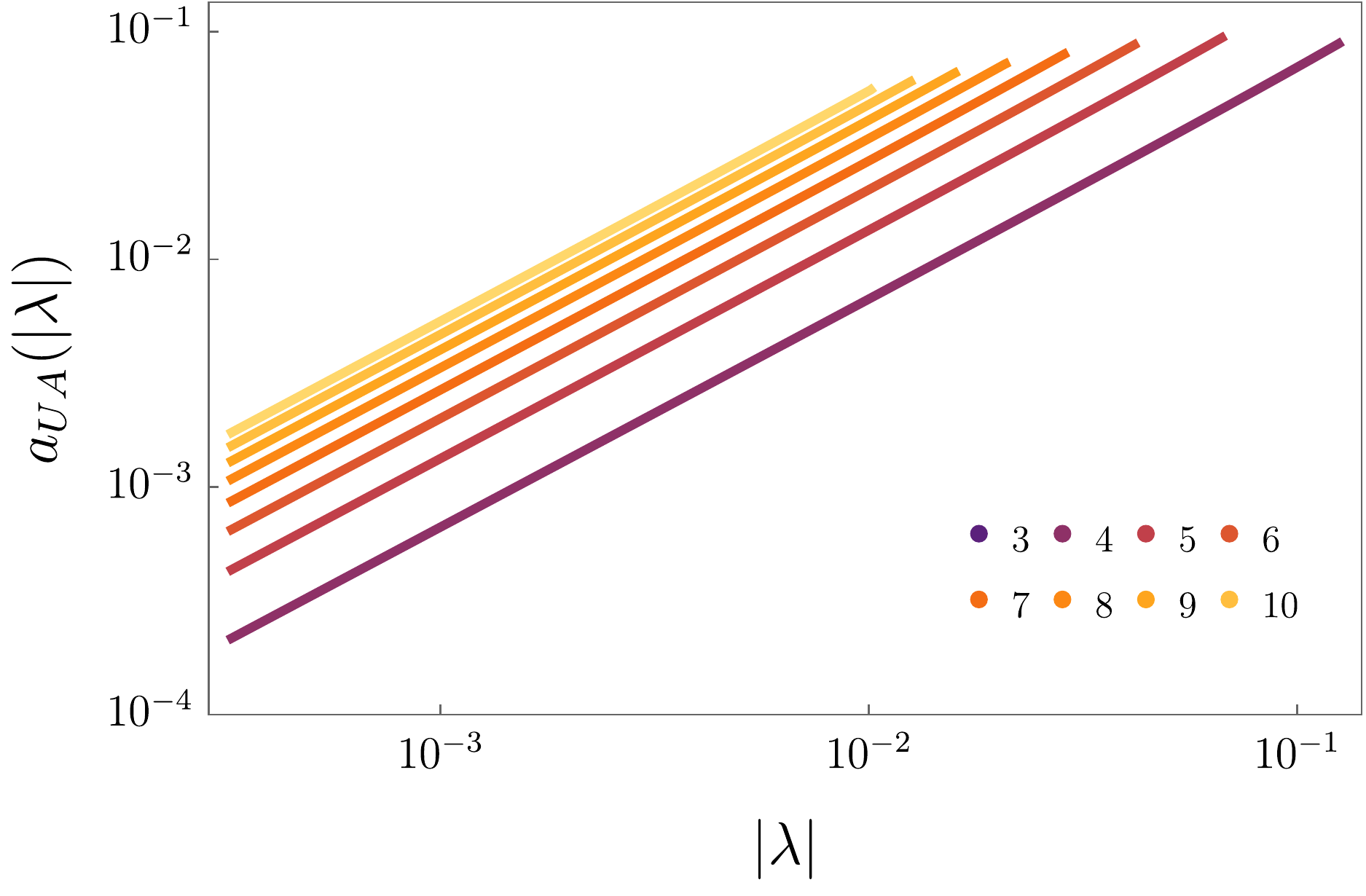}}	
\caption{Plot (log-log) of (a) the EA and (b) the UA asymmetry ratios, \eqq{asyBothDC}, as a function of the absolute value of the depolarizing parameter, $\abs{\lambda}$, for several system dimensions, $d=2,3,\cdots,10$ (from darker to lighter colours). Note that the UA asymmetry starts from $d=3$ since it is zero for $d=2$. The general feature is an increase of the asymmetry ratio with both $\abs{\lambda}$ and $d$. See the text for further comments.}
\label{fig2}
\end{figure}

\section{Reciprocal pairs} \label{SECREC}

The results presented in the previous section can be framed in a broader context by considering
pairs of quantum channels $\mathbf{D}_{\pm}$ producing 
output images  that  are geometrically equivalent, in the sense that one
can pass from the output of a channel to its reciprocal via  
 a rigid transformation $\Lambda$ and vice versa, i.e.,
 \begin{eqnarray} 
 \Lambda \circ \mathbf{D}_{+} = \mathbf{D}_{-} \;, \quad  \Lambda^{-1}  \circ \mathbf{D}_{-} = \mathbf{D}_{+} \label{CENTRAL111} \;,\end{eqnarray} 
 with 
\begin{eqnarray}
 \| \Lambda \left(\hat{\rho}_1\right)-\Lambda\left(\hat{\rho}_2\right)\| &=&  \label{DISTalsoAA} 
  \| \hat{\rho}_1-\hat{\rho}_2\|\;, 
 \end{eqnarray} 
 for all $\hat{\rho}_1$, $\hat{\rho}_2 \in \mathfrak{S} ({\cal H})$. 
In a classical setting this property would be sufficient to make $\mathbf{D}_{\pm}$  equivalent in terms of communication efficiency. 
The same happens also in the quantum case   when $\Lambda$ can be identified with a unitary transformation since then 
 one can invoke the data-processing inequality to show that $\mathbf{D}_{\pm}$ possess identical communication capacities. 
 However, if $\Lambda$ is not  completely positive, then the  above argument is no longer applicable and asymmetric behavior may arise. A first example of this fact is provided by 
the reciprocal  DC pairs $\{ \mathcal{D}_{\pm |\lambda|}^{(d)}\}$ of Sec.~\ref{sec:qubitDep} whose connecting map is
realized by the inversion with respect to the completely mixed point, i.e., the (not completely-positive) super-operator 
 \begin{eqnarray}  \label{LAMBDA} 
 \Lambda = \Lambda^{-1} : = - {\rm Id} + 2 {\cal D}_0^{(d)}\;.
 \end{eqnarray}  
 Other instances of the same effect can be obtained by considering the pairs 
 $\{ \mathbf{D}^{(\Phi)}_{\pm |\lambda|}\}$, where 
$\mathbf{D}^{(\Phi)}_{\lambda}$ is obtained by linearly combining the output of an assigned quantum channel $\Phi$ with the fully depolarizing map ${\cal D}_0^{(d)}$,
\begin{equation}\label{dcGEN}
\mathbf{D}^{(\Phi)}_{\lambda}=\lambda \Phi+\left(1-\lambda\right){\cal D}_0^{(d)},
\end{equation} 
and $\lambda$ is properly chosen to ensure the complete positivity.
Similarly to the DC case, recovered from~(\ref{dcGEN}) by setting $\Phi= {\rm Id}$, the channels $\mathbf{D}^{(\Phi)}_{|\lambda|}$ and $\mathbf{D}^{(\Phi)}_{-|\lambda|}$ act on $\mathfrak{S}({\cal H})$ 
as uniform contractions, mapping the same input $\hat{\rho}$ into two distinct output configurations that are equally distant from the completely mixed state but on opposite sides of it, i.e.,
\begin{eqnarray} 
  \| \mathbf{D}_{\pm |\lambda|}^{(\Phi)}\left(\hat{\rho}_1\right)-\mathbf{D}_{\pm |\lambda|}^{(\Phi)}\left(\hat{\rho}_2\right)\| &=&  \label{DISTalsoII} 
|\lambda|\;  \| \Phi(\hat{\rho}_1)-\Phi(\hat{\rho}_2)\|\;, \\
\|\mathbf{D}^{(\Phi)}_{|\lambda|}\left(\hat{\rho}\right)- \mathbf{D}^{(\Phi)}_{-|\lambda|}\left(\hat{\rho}\right)\| &=& \|\mathbf{D}^{(\Phi)}_{\pm|\lambda|}\left(\hat{\rho}\right)- \frac{\hat{\mathbf{1}}}{d}\|\;,  \label{DIST111}  
\end{eqnarray}
for all $\hat{\rho}, \hat{\rho}_1, \hat{\rho}_2 \in \mathfrak{S}({\cal H})$. As before, Eq.~(\ref{DISTalsoII}) holds also when the channels act locally on extended states of the system S with an ancillary system A. 
Accordingly, the pair $\{ \mathbf{D}^{(\Phi)}_{\pm |\lambda|}\}$ produces output images of $\mathfrak{S} ({\cal H})$ which are connected as in (\ref{CENTRAL111}) via the super-operator~(\ref{LAMBDA}).
Our previous observations then suggest that, also in this general case,  the two reciprocal channels should have an asymmetric performance in terms of communication efficiency. 
Unfortunately,  for an arbitrary choice of $\Phi$ this conjecture  
is not easy to confirm due to the difficulty in 
computing  the exact value of the capacities required to evaluate the ratios 
\begin{equation}\label{asymWccNEW}
a_{\ell}\left(\abs{\lambda}\right)=\frac{C_{\ell}\left(\mathbf{D}^{(\Phi)}_{-|\lambda|}\right)-C_{\ell}\left(\mathbf{D}^{(\Phi)}_{|\lambda|}\right)}{C_{\ell}\left(\mathbf{D}^{(\Phi)}_{|\lambda|}\right)}\;.
\end{equation}
There  are cases however where this can be done, e.g., when $\Phi$ is a Weyl-covariant map
~\cite{holWeylCov,holevoRemarksFull,holDattaWCC}
of arbitrary dimension. 
Before presenting these results, let us first comment on some general properties of the reciprocal channel pairs we have just introduced:
\begin{enumerate}
\item  The map  $\mathbf{D}^{(\Phi)}_{\lambda}$ defined in Eq.~(\ref{dcGEN})   cannot be seen as a simple concatenation of the DC $\mathcal{D}_{\lambda}^{(d)}$  with 
$\Phi$. Indeed,  due to the fact that the output states of $\Phi$ form in general a proper subset of all possible density matrices of the system $S$, in Eq.~(\ref{dcGEN}) the parameter $\lambda$ can take
values on an interval ${\cal I}(\Phi)$ which is broader than the one required by  the complete positivity of  $\mathcal{D}_{ \lambda}^{(d)}$, i.e., $[\lambda_{m}(d),1]\subseteq{\cal I}(\Phi)$ -- see below for an explicit example;
\item For any assigned $\Phi$ the quantum channels  (\ref{dcGEN})  form a convex set, i.e., given   $\lambda_1,\lambda_2\in{ \cal I}(\Phi)$ we have 
\begin{eqnarray} 
\label{CONVEX111} 
p \mathbf{D}^{(\Phi)}_{\lambda_1} +  (1-p) \mathbf{D}^{(\Phi)}_{\lambda_2} &=&\mathbf{D}^{(\Phi)}_{\overline{\lambda}}\;,  
\end{eqnarray} 
  with $\overline{\lambda}=p \lambda_1 + (1-p) \lambda_2\in { \cal I}(\Phi)$ for all $ \forall p\in[0,1]$.
A generalization of~(\ref{SEMIG}) can also be provided in the following form 
\begin{eqnarray} 
\mathcal{D}_{\lambda}^{(d)}\circ \mathbf{D}_{\lambda_1}^{(\Phi)} &=&\mathbf{D}_{\lambda \lambda_1}^{(\Phi)}\;, \label{SEMIG11}
\end{eqnarray} 
which holds for all $\lambda\in [ \lambda_m(d),1]$ and $\lambda_1 \in {\cal I}(\Phi)$;
\item Equations~(\ref{CONVEX111}) and~(\ref{SEMIG11}) induce a natural ordering on the capacities of the maps $\mathbf{D}^{(\Phi)}_{\lambda}$, as did the convex semigroup properties in the DC case, see Eq.~(\ref{AASS}).
Indeed, while Eq.~(\ref{CONVEX111}) implies that $C_{\ell}\left(\mathbf{D}_{\lambda}^{(\Phi)}\right)$ is a convex function of the parameter $\lambda$, Eq.~(\ref{SEMIG11}) implies that $C_{\ell}\left(\mathbf{D}_{\lambda_2}^{(\Phi)}\right) \leq 
  C_\ell \left(\mathbf{D}_{\lambda_1}^{(\Phi)}\right)$ for all $\lambda_1,\lambda_2 \in {\cal I}(\Phi)$ admitting a $\lambda\in [ \lambda_m(d),1]$  such
that $\lambda_2 = \lambda \lambda_1$. 
  It is worth stressing however that none of these properties can be used to establish a hierarchy between the elements of a given reciprocal pair  $\{\mathbf{D}^{(\Phi)}_{\pm|\lambda|}\}$;
\item
If   $\Phi$ is the DC $\mathcal{D}_{\lambda_1}^{(d)}$, then the associated 
$\mathbf{D}^{(\Phi)}_{\lambda}$ is also a DC with  depolarizing parameter $\lambda \lambda_1$, i.e.,
$\mathbf{D}^{(\Phi)}_{\lambda} = \mathcal{D}_{\lambda \lambda_1}^{(d)}$. 
Accordingly, complete positivity of the reciprocal pairs  $\{\mathbf{D}^{(\Phi)}_{\pm|\lambda|}\}$ is ensured if and only if $|\lambda| \leq |\lambda_m(d)|/|\lambda_1|$
 and  the presence of 
asymmetries then trivially follows from that of a simple DC by fixing $\lambda_{1}$. Interestingly, the sign of the corresponding asymmetry ratio can be flipped, i.e.,
\begin{equation}\label{asyBothDC111}
a_{\ell}(\abs{\lambda})=\frac{C_{\ell}\left(\mathcal{D}_{-\abs{\lambda}\lambda_1}^{(d)}\right)-C_{\ell}\left(\mathcal{D}_{\abs{\lambda}\lambda_1}^{(d)}\right)}{C_{\ell}\left(\mathcal{D}_{\abs{\lambda}\lambda_1}^{(d)}\right)},
\end{equation}
becomes negative for $\lambda_{1}<0$.
 Similarly, if a channel $\mathbf{D}^{(\Phi)}_{\lambda}$ is found to exhibit asymmetric capacity with positive asymmetry ratio, one can always build another channel 
 \begin{eqnarray}
 \mathbf{D}^{(\mathbf{D}^{(\Phi)}_{\lambda_1})}_{\lambda}&=&\mathbf{D}^{(\Phi)}_{\lambda\lambda_1}\;,
\label{dcGEN123121223}
\end{eqnarray}
which again exhibits asymmetry but with opposite sign of the ratio.
\end{enumerate}

\section{Weyl-covariant reciprocal pairs} 
\label{sec:weylCov}

Weyl-covariant channels~\cite{holWeylCov,holevoRemarksFull,holDattaWCC} can 
 be defined in any finite dimension by introducing the generalization of Pauli matrices, i.e., finite-dimensional Weyl operators:
\begin{equation}\label{wDef}
\hat{W}_{z}=\hat{W}_{(x,y)}=\hat{U}^{x}\hat{V}^{y}\quad\forall x,y\in\mathbb{Z}_{d},
\end{equation}
where $\mathbb{Z}_{d}$ is the additive cyclic group in dimension $d$, whose elements are labelled by $\{0,\cdots,d-1\}$, and $z\in Z=\mathbb{Z}_{d}\oplus\mathbb{Z}_{d}$. The operators in \eqq{wDef} are unitary: $\hat{U}$ operates on a given basis $\{\ket{e_{j}}\}_{j=1}^{d}$ of the Hilbert space of the system by increasing the state index, while $\hat{V}$ by adding a phase factor, i.e.,
\begin{equation}\label{wDef2}
\hat{U}\ket{e_{j}}=\ket{e_{(j+1)\text{mod} d}},\qquad\hat{V}\ket{e_{j}}=e^{\frac{2i\pi}{d} j}\ket{e_{j}}.
\end{equation}
The transformations~(\ref{wDef})  constitute an irreducible unitary representation $W(Z)$ of $Z$ on the space of operators of the system. They also are the finite-dimensional counterpart to displacement operators of continuous-variables systems; in particular, they obey similar canonical commutation relations~\cite{holDattaWCC,serafiniBOOK}, the orthogonality condition
\begin{equation}\label{ortho}
\tr{}{\hat{W}_{z}\hat{W}_{z'}^{\dag}}=d \delta_{z,z'}
\end{equation}
and the completeness relation
\begin{equation}\label{complete}
\frac{1}{d^{2}}\sum_{z}\hat{W}_{z}\hat{X}\hat{W}_{z}^{\dag}=\tr{}{\hat{X}}\frac{\hat{\mathbf{1}}}{d},
\end{equation}
for any operator $\hat{X}$ on the Hilbert space of the system.\\
A quantum channel $\Phi$ is said to be a Weyl-covariant channel (WCC) if it satisfies
\begin{equation}
\Phi\left(\hat{W}_{z}\hat{X}\hat{W}_{z}^{\dag}\right)=\hat{W}_{z}\Phi\left(\hat{X}\right)\hat{W}_{z}^{\dag}\quad\forall z\in Z,
\end{equation}
i.e., if its action commutes with that of any Weyl operator. 
By construction WCCs form a convex set and admit a canonical representation as convex combinations of Weyl unitary transformations~\cite{holDattaWCC}.
Accordingly, the most generic WCC can always be written as 
\begin{equation}\label{wccExp}
\Phi_{\{p_{z}\}}\left(\hat{\rho}\right)=\sum_{z}p_{z}\hat{W}_{z}\hat{\rho}\hat{W}_{z}^{\dag},
\end{equation}
with $\{p_{z}\}$ a probability distribution on $Z$. Eq.~(\ref{wccExp}) makes it explicit that WCCs form a proper subset of random unitary channels and of unital channels. In particular, in dimension two, the Weyl operators are just the ordinary Pauli matrices (up to an irrelevant phase on $\sigma_{2}$) and qubit WCCs coincide with the whole class of unital qubit channels up to unitary operations~\cite{kingRuskaiUnital}. 
It is also worth observing that 
the DC defined in \eqq{dc} is a special instance of WCC, as can be shown by applying \eqq{complete} to the second term in the sum and observing that $\hat{W}_{(0,0)}=\hat{\mathbf{1}}$. Accordingly we can express $\mathcal{D}_{\lambda}^{(d)}$ in the form 
\eqq{wccExp}, 
by taking $p_{z}(\lambda)$ equal to $\frac{1+(d^{2}-1)\lambda}{d^{2}}$ for $z=(0,0)$ and $\frac{1-\lambda}{d^{2}}$ otherwise. In Appendix~\ref{WCCQM} we also show that WCCs can reach  the same input-output fidelity threshold attained by DCs when averaged over the set of pure input states.

A close inspection of Eq.~(\ref{dcGEN}) reveals that the  maps $\mathbf{D}^{(\Phi)}_{\lambda}$   associated  to  a WCC $\Phi_{\{q_z\}}$
 are also Weyl-covariant and can be hence expressed in the canonical form~(\ref{wccExp}) with a proper choice of the probability distribution $p_z$.  Specifically this is 
\begin{equation}\label{chVars}
p_{z}=\frac{1+(d^{2}q_{z}-1)\lambda}{d^{2}}\quad\forall z\in Z(d),
\end{equation}
where 
 $\{ q_z\}$ is the probability distribution associated with the canonical form of the input WCC  $\Phi_{\{q_z\}}$. Equation~(\ref{chVars}) can be used to determine the range  of values of the parameter $\lambda$ 
 that ensures complete positivity of the transformations $\mathbf{D}^{(\Phi)}_{\lambda}$ -- an explicit proof of this is provided in Appendix~\ref{app:cp}. Specifically, by imposing $p_z$ to take values in the interval $[0,1]$ we get the following constraint 
\begin{equation}\label{range}
\lambda\in[\lambda_{m}(\{q_{z}\}),\lambda_{M}(\{q_{z}\})],
\end{equation}
with $\lambda_{m}(\{q_{z}\})$ and $\lambda_{M}(\{q_{z}\})$ being negative, resp. positive, quantities defined by the identities 
\begin{equation}
\begin{aligned}
&\lambda_{m}\left(\{q_{z}\}\right)=\max_{q_{z}>d^{-2}}\frac{1}{1-d^{2}q_{z}},\\
&\lambda_{M}\left(\{q_{z}\}\right)=\min_{q_{z}<d^{-2}}\frac{1}{1-d^{2}q_{z}}.
\end{aligned}
\end{equation}
Therefore, given a generic WCC of canonical form $\Phi_{\{q_{z}\}}$ we can now construct reciprocal pairs $\mathbf{D}^{(\Phi_{\{q_{z}\}})}_{\pm|\lambda|}$ 
for all $|\lambda|$ satisfying the constraint 
\begin{eqnarray}\label{NEWRANGE}  \abs{\lambda}\leq\min\{\abs{\lambda_{m}(\{q_{z}\})},\abs{\lambda_{M}(\{q_{z}\}})\}\;.\end{eqnarray} 

We are now ready to study the presence of asymmetry in the reciprocal pairs $\mathbf{D}^{(\Phi_{\{q_{z}\}})}_{\pm|\lambda|}$. 
Let us start by noting that neither the UA nor the EA classical capacities of WCCs are known, apart from the case $d=2$ where they coincide with unital qubit channels~\cite{kingUnital}. Indeed it has been shown in Ref.~\cite{cortese,holevoRemarks,holevoRemarksFull} that the UA capacity of these channels can be computed by minimizing the output entropy, but still there is no proof that the additivity holds in $d>2$, see~\cite{kingUnital,depolarizingCCap}, so that the problem may be hard to solve. Luckily, the EA capacity is always additive and in the following we explicitly compute its value. We remind that  the EA capacity~\cite{bennetEntAss2} of a channel $\Phi$ is provided by the following expression 
\begin{equation}\label{defEA}
C_{EA}(\Phi)=\max_{\hat{\rho}}I\left(\hat{\rho},\Phi\right)\end{equation}
where $I\left(\hat{\rho},\Phi\right)$ is the quantum mutual information~\cite{nChuangBOOK} of the channel $\Phi$ with input state $\hat{\rho}$. For the class of WCCs we have:
\begin{eqnarray}
&I\big(\hat{\rho}&,\Phi_{\{p_{z}\}}\big)=\ave{I\left(\hat{\rho},\mathcal{W}\circ\Phi_{\{p_{z}\}}\right)}_{\mathcal{W}\in W(Z)}\nonumber\\
&&=\ave{I\left(\mathcal{W}\left(\hat{\rho}\right),\Phi_{\{p_{z}\}}\right)}_{\mathcal{W}\in W(Z)}\\
&&\leq I\left(\ave{\mathcal{W}\left(\hat{\rho}\right)}_{\mathcal{W}\in W(Z)},\Phi_{\{p_{z}\}}\right)=I\left(\frac{\hat{\mathbf{1}}}{d},\Phi_{\{p_{z}\}}\right)\nonumber,
\end{eqnarray}
where the first equality follows from the unitary-invariance of the quantum mutual information, $\mathcal{W}(\cdot)=\hat{W}\cdot\hat{W}^{\dag}$ is the unitary map associated to a Weyl operator $\hat{W}$ and we have introduced the average $\ave{\cdot}$ over the Weyl group $W(Z)$. The second equality instead follows from the Weyl-covariance of WCCs, while the inequality is due to the concavity of quantum mutual information as a function of the input state. Finally, the last equality follows from the completeness relation of \eqq{complete}. Hence the maximum in \eqq{defEA} is attained by the maximally mixed state for any WCC $\Phi_{\{p_{z}\}}$. The computation of its value is reported in Appendix~\ref{app:cp} and results in:
\begin{equation}\label{cEaWcc}
C_{EA}\left(\Phi_{\{p_{z}\}}\right)=2\log d-H(\{p_{z}\}).
\end{equation}
\begin{figure}[t!]
\includegraphics[scale=.44]{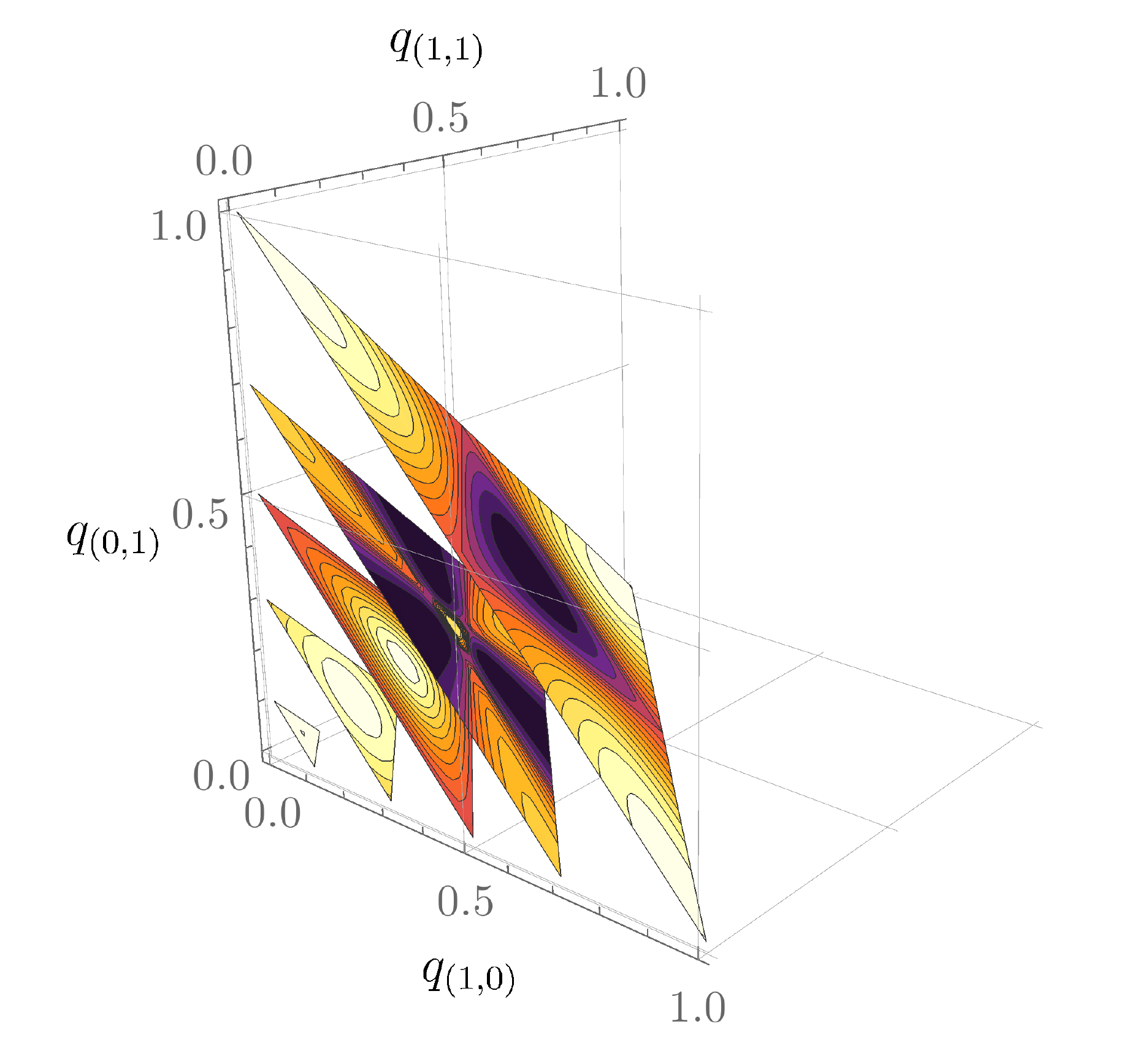}\put(-60,80){\includegraphics[scale=.35]{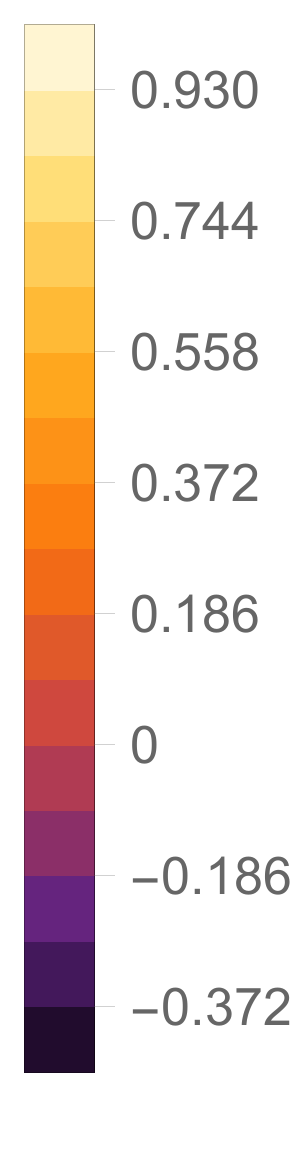}}
\put(-60,200){\includegraphics[scale=.35]{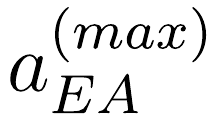}}	
\caption{3D slice-plot of the maximum asymmetry ratio (\ref{asymWccNEW})  of the EA capacity of WCC-based reciprocal pairs, $\mathbf{D}_{\pm\abs{\lambda}}^{(\Phi_{\{q_{z}\}})}$, in the space determined by the three probability values $q_{z}$ for $z\neq(0,0)$ in $d=2$. The brighter regions are those where the asymmetry ratio is larger, corresponding to the DC (all three probabilities equal) and to its equivalent channels obtained by the application of a Weyl unitary. Darker regions see instead a negative asymmetry ratio, hence a gain of the non-inverting channel of the pair, as discussed in Sec.~\ref{SECREC}.}
\label{fig3}
\end{figure}
Strikingly, this formula says that the EA capacity of a noisy WCC channel is that of the ideal channel, $2\log d$, minus the Shannon entropy of the probability distribution that determines the weights of each Weyl map in the composition of the channel. Expressing now the reciprocal pairs $\mathbf{D}^{(\Phi_{\{q_{z}\}})}_{\pm|\lambda|}$ of a given WCC $\Phi_{\{q_{z}\}}$ in the canonical form~(\ref{wccExp})  with probabilities
(\ref{chVars}) we get 
\begin{equation}
C_{EA}\left(\mathbf{D}^{(\Phi_{\{q_{z}\}})}_{\pm|\lambda|}\right)=2\log d-H\left(\left\{\frac{1\pm (d^{2}q_{z}-1)|\lambda|}{d^{2}}\right\}\right)\;,
\end{equation}
for all $|\lambda|$ as in Eq.~(\ref{NEWRANGE}).
A numerical study of the associated asymmetry ratio (\ref{asymWccNEW}) 
 reveals that for all assigned choices of $\{q_{z}\}$  it reaches its maximum for 
 the highest allowed value of $\abs{\lambda}$ (please notice that this feature cannot be directly associated with the 
 general properties of $C_{\ell}\left(\mathbf{D}_{\lambda}^{(\Phi)}\right)$ discussed in Sec.~\ref{SECREC}). 
  Moreover, a maximization over  the probability distribution $\{q_{z}\}$ of the input WCC reveals that the DC exhibits the largest asymmetry ratio among all such channels and it is unique up to Weyl-unitary transformations. However, there is a whole range of channel parameters $\{q_{z}\}$ where a high asymmetry ratio can be attained, e.g., see Fig.~\ref{fig3} for the case of $d=2$. The UA capacity in this latter case is known~\cite{kingUnital} and symmetric in $\lambda$, as for the DC.

\section{Discussion}\label{sec:conc}
Inspired by the results of Refs.~\cite{peresFlip,massarFlip,gisinFlip,munozTapiaFlip,massarFlip2} we discuss  the communication efficiency of pairs of
channels which produce output images of the input signals that are geometrically equivalent,  i.e., they are connected via a rigid transformation which allows one to reconstruct locally the outcome of one map from
that of the other one. At variance with what would happen in a classical scenario, an asymmetric behaviour can arise, yielding different capacity values to the two elements of the pair. 
In our analysis we focused on the special case of classical information transfer, evaluating, when possible, the unassisted, $C_{UA}$, and entanglement-assisted, $C_{EA}$, capacities of the channels.
Analogous behaviours are expected however also for the transferring of quantum messages, if the quantum capacity is non-zero (in the case of the entanglement-assisted quantum capacity $Q_{EA}=C_{EA}/2$ this is just a trivial consequence
of our finding). 

\section{Acknowledgements}
The authors would like to thank Andrea Mari for many useful discussions on the topics treated in the article. M.R.~acknowledges funding by the Spanish MINECO (projects FIS2013-40627-P and FIS2016-80681-P (AEI/FEDER, UE)).

\appendix
\section{DCs as optimal quantum movers} \label{movers} 

Quantum movers (QMs) were defined in Ref.~\cite{qMovers} as quantum channels that move any pure state of the system by a constant amount in phase-space. Specifically, taking the 
input-output fidelity~\cite{nChuangBOOK} as a measure of the phase-space overlap between the pure input state and the associated output state,
a  QM  $\Phi$ satisfies the following constraint:
\begin{equation}\label{qMoverDef}
\bra{\psi}\Phi\left(\ketbra{\psi}\right)\ket{\psi}=f(\Phi)\quad\forall \ket{\psi}\text{ pure},
\end{equation}
where $f(\Phi)$ is a quantity that depends only on the channel parameters but is constant otherwise. 
A trivial QM is the identity channel $\rm{Id}$, which does not move input states at all, i.e., $f(\rm{Id})=1$.
 QMs with lower values of $f(\Phi)$ move the input states progressively further away in phase-space and a QM attaining null $f(\Phi)$   would virtually turn any pure input state into its orthogonal. However, it is well-known that the existence of a universal inverter is not allowed by quantum mechanics~\cite{uNot1,uNot2,uNot3} and one can at best realize an approximate inverter. The class of QMs thus comprises approximate inverters that allow to tune the input-output overlap to any physically-allowed value. 
The DC maps ${\cal D}_\lambda^{(d)}$  defined in (\ref{dc}) are QMs with a moving parameter that
linearly depends upon $\lambda$ as indicated in Eq.~(\ref{MOVERS}). 
They are  optimal in the sense that, by playing with $\lambda$, we can span the full spectrum of the allowed values of the moving parameter $f(\Phi)$: a proof of this fact is presented here closing the problem left open in~\cite{qMovers}.
To see this explicitly  note that 
\eqq{qMoverDef} can be equivalently expressed as 
\begin{equation}
\bra{0}\hat{U}^{\dag}\Phi\left(\hat{U}\ketbra{0}\hat{U}^{\dag}\right)\hat{U}\ket{0}=f\left(\Phi\right)\quad\forall \hat{U}\in U(d).
\end{equation}
where we wrote $\ket{\psi}$ as $\hat{U}\ket{0}$ with $\hat{U}$ being a generic element of the  unitary group  $U(d)$ and $\ket{0}$ being an (arbitrary) reference state. We are interested in minimizing the function $f(\Phi)$ over all channels that satisfy the QM condition, \eqq{qMoverDef}.
Taking the average with respect to $\hat{U}$ through the Haar measure $d\mu(\hat{U})$ of $U(d)$  we can then identify a new QM, $\overline\Phi = \int d\mu(\hat{U}) \hat{U}^{\dag}\Phi\left(\hat{U}\cdots\hat{U}^{\dag}\right)\hat{U}$, which is characterized by the same moving strength  of the original one, i.e.,
$f(\overline{\Phi}) = f(\Phi)$, and which commutes with the unitary group. 
 This last is a very stringent requirement that is satisfied only by the DCs, as detailed in Ref.~\cite{uNot3} using the methods of~\cite{twirling1}. 
 Accordingly we can conclude that 
 \begin{equation}\label{opQMover}
\min_{\Phi}f\left(\Phi\right)=\min_{\bar{\Phi}}f\left(\bar{\Phi}\right)=\min_{\lambda}f({\lambda})=f(\lambda_{m}(d))=\frac{1}{d+1}.
\end{equation}

\section{Weak quantum movers and WCCs}\label{WCCQM} 
The WCCs introduced in Sec.~\ref{sec:weylCov} do not satisfy the quantum mover condition~(\ref{qMoverDef}). Yet, as we shall see in the following, 
when  averaging over all possible pure input states,  they are capable of reaching the same input-output fidelity threshold (\ref{opQMover})  achieved by optimal quantum movers. 
To see this, given $\Phi$ a generic quantum channel  let us consider 
the quantity 
\begin{equation} \label{AIOF} 
\bar{f}\left(\Phi\right)=\int d \mu({\psi})\;  \langle \psi|  \Phi\left(\ketbra{\psi}\right)\ket{\psi}\;,
\end{equation}
where the integral is performed with the Haar measure $d \mu({\psi})$ over pure states.  Note that this quantity can be defined for all channels, not only those satisfying the stronger QM condition, \eqq{qMoverDef}. In particular, in the latter case it holds trivially $\bar{f}(\Phi_{QM})=f(\Phi_{QM})$. Moreover, its minimum value over all quantum channels coincides with the one attained by QMs, since $\bar{f}(\Phi)=f(\bar{\Phi})$ and we can apply \eqq{opQMover}.For a generic channel instead we can compute this quantity using the twirling method~\cite{twirling1,twirlAnto} as follows:
\begin{eqnarray}
&&\bar{f}\left(\Phi\right)=\int dU\sum_{k}\bra{0}\hat{U}^{\dag}\hat{M}_{k}\hat{U}\ket{0}\bra{0}\hat{U}^{\dag}\hat{M}_{k}^{\dag}\hat{U}\ket{0}\nonumber\\
&&=\int dU\sum_{k}\bra{00}\left(\hat{U}^{\dag}\otimes\hat{U}^{\dag}\right)\left(\hat{M}_{k}\otimes\hat{M}_{k}^{\dag}\right)\left(\hat{U}\otimes\hat{U}\right)\ket{00}\nonumber\\
&&=\sum_{k}\bra{00}\mathcal{T}\left(\hat{M}_{k}\otimes\hat{M}_{k}^{\dag}\right)\ket{00},
\end{eqnarray}
where the first equality follows from writing the integral as an average over the unitary group and employing a Kraus representation~\cite{nChuangBOOK} for the channel, i.e., $\Phi(\cdot)=\sum_{k}\hat{M}_{k}\cdot\hat{M}_{k}^{\dag}$, the second one from introducing a trivial copy of the system and the third one by defining the twirling map $\mathcal{T}$ as the average over the $\hat{U}^{\dag}\otimes\hat{U}^{\dag}$ representation of the unitary group~\cite{twirling1}. Its action is to project on the commutant of the corresponding group, which is spanned by the identity, $\hat{\mathbf{1}}^{\otimes 2}$, and the swap operator, $\hat{S}=\sum_{i,j=1}^{d}\dketbra{ij}{ji}$, see~\cite{twirling1,twirlAnto}. The average~(\ref{AIOF}) then becomes
\begin{eqnarray}\label{QUESTA} 
\bar{f}\left(\Phi\right)&=&\frac{1}{d(d+1)}\sum_{k}\tr{}{\left(\hat{M}_{k}\otimes\hat{M}_{k}^{\dag}\right)\left(\hat{\mathbf{1}}^{\otimes 2}+\hat{S}\right)}\nonumber\\
&=&\frac{\sum_{i=1}^{r}\nu_{i}+d}{d(d+1)},\end{eqnarray}
where we have employed the definition of the swap operator and $\{\nu_{i}\}_{i=1}^{r}$, $r\leq d^{2}$, are the eigenvalues of the map $\Phi$.
In particular, for the class of WCCs just presented, Eq.~(\ref{QUESTA}) can be simplified by using the canonical form of \eqq{wccExp} to identify  $\hat{M}_{z}$ with the operators $\sqrt{p_{z}}\hat{W}_{z}$. Hence from  Eqs.~(\ref{wDef} -- \ref{ortho}), it follows that 
\begin{equation}\label{aiof}
\bar{f}\left(\Phi_{\{p_{z}\}}\right)=\frac{d p_{(0,0)}+1}{d+1} \geq \frac{1}{d+1}\;, 
\end{equation}
the inequality being saturated by the WCCs  (\ref{wccExp}) that have zero identity component, i.e., $p_{(0,0)}=0$. 

\section{Complete positivity and EA capacity of WCCs}\label{app:cp}
In this Appendix we analyze the Choi-Jamiolkowski state of WCCs, useful both for defining the allowed range of values of $\lambda$, \eqq{range}, and computing the EA capacity, \eqq{cEaWcc}.
The complete-positivity condition on a quantum channel $\Phi$ can be imposed by requiring that its Choi-Jamiolkowsi (CJ) state is positive semidefinite~\cite{choiCP}. The latter is obtained by applying an extended version of the channel, $\rm{Id}\otimes\Phi$, to the maximally entangled state $\ket{\Omega}$. Moreover, the quantum mutual information~\cite{wildeBOOK} employed in \eqq{defEA} is defined, mimicking its classical counterpart, as the sum of the input and output entropies of the channel, minus the entropy of its extended version, i.e.,
\begin{equation}\label{qMInfo}
I(\hat{\rho},\Phi)=S(\hat{\rho})+S(\Phi(\hat{\rho}))-S((\rm{Id}\otimes\Phi)(\ketbra{\rho})),
\end{equation}
where $\ketbra{\rho}$ is a purification of the input state of the system. For the particular case of WCCs the optimal state to insert in \eqq{qMInfo} is $\hat{\rho}=\hat{\mathbf{1}}/d$, as shown in Sec.~\ref{sec:weylCov}. Hence the first and second terms in \eqq{qMInfo} are maximum and equal to $\log d$, since WCCs are unital, while the third term amounts to the entropy of the CJ state of the channel, since the purification of the maximally mixed state is $\ket{\Omega}$ itself. \\
Both the complete-positivity condition and the EA capacity of WCCs thus depend on the spectrum of the CJ state $(\rm{Id}\otimes\Phi_{\{p_{z}\}})\ketbra{\Omega}$. The latter can be more easily computed by considering the complementary channel: 
\begin{equation}\label{comple}
\tilde{\Phi}_{\{p_{z}\}}(\hat{\rho})=\tr{S}{\hat{U}_{SE}\left(\hat{\rho}_{S}\otimes\ketbra{\psi}_{E}\right)\hat{U}_{SE}^{\dag}},
\end{equation}
where we have introduced the Stinespring representation~\cite{stinespring} of $\Phi_{\{p_{z}\}}$ by coupling the system, $\hat{\rho}_{S}$, to an environment,
 \begin{equation}\label{stateSt}
\ket{\psi}_{E}=\sum_{z}\sqrt{p_{z}}\ket{z}_{E},
\end{equation}
 through a unitary interaction,
 \begin{equation}\label{uniSt}
 \hat{U}_{SE}=\sum_{z}(\hat{W}_{z})_{S}\otimes\ketbra{z}_{E}.
 \end{equation}
Here $\{\ket{z}\}_{z\in Z}$ is a $d^{2}$-dimensional orthonormal basis of system E. It can be checked that tracing out the environment instead of the system in \eqq{comple} returns the WCC of \eqq{wccExp}. Since the total output state comprising the system, its purifying counterpart and the environment is pure, the local spectra (and hence the entropies) of any of its bipartition are equal and we can write:
\begin{equation}
S\left(\left(\rm{Id}\otimes\Phi_{\{p_{z}\}}\right)(\ketbra{\Omega})\right)=S\left(\tilde{\Phi}_{\{p_{z}\}}\left(\frac{\hat{\mathbf{1}}}{d}\right)\right).
\end{equation}
The latter term now is easy to compute thanks to the properties of Weyl-unitary operators. Indeed we have:
\begin{equation}
\begin{aligned}
\tilde{\Phi}_{\{p_{z}\}}\left(\frac{\hat{\mathbf{1}}}{d}\right)&=\sum_{z,z'}\frac{1}{d}\tr{}{\hat{W}_{z}\hat{W}_{z'}^{\dag}}\sqrt{p_{z}p_{z'}}\dketbra{z}{z'}\\
&=\sum_{z}p_{z}\ketbra{z},
\end{aligned}
\end{equation}
where we have substituted the expressions of Eqs.~\eqref{uniSt},\eqref{stateSt} in \eqq{comple} for $\Phi_{\{p_{z}\}}$ and employed the orthogonality condition \eqq{ortho}. Therefore we conclude that the spectrum of the CJ state of a WCC is simply the probability distribution of its Weyl-unitary components. From this it directly follows the expression \eqq{cEaWcc} of the EA capacity, while the complete-positivity condition just amounts to require that $\{p_{z}\}$ is a probability distribution.

\end{document}